\documentclass[12pt,epsfig]{article}
\usepackage{amsmath}
\usepackage{graphicx}
\usepackage{epstopdf}
\usepackage{titlesec}
\usepackage[usenames, dvipsnames]{color}
\titleformat*{\section}{\large\bfseries\sffamily}
\titleformat*{\subsection}{\small\bfseries\sffamily}
\begin{document}
\begin{center}
A new nonparametric test for two sample multivariate location problem with application to astronomy\\
\vspace{.1 in}
Soumita Modak$^{1}$  and Uttam Bandyopadhyay$^2$ \\
Department of Statistics, Calcutta University, India\\
email: soumitamodak2013@gmail.com$^1$,  ubandyopadhyay08@gmail.com$^2$
\end{center}
\begin{abstract}
This paper provides a nonparametric test for the identity of two multivariate continuous distribution functions (d.f.'s) when they differ in locations. The test uses Wilcoxon rank-sum statistics on distances between observations for each of the components and is unaffected by outliers. It is numerically compared with two existing procedures in terms of power. The simulation study shows that its power is strictly increasing in the sample sizes and/or in the number of components. The applicability of this test is demonstrated by use of two astronomical data sets on early-type galaxies.
\end{abstract}
keywords: {nonparametric multivariate test; combination of Wilcoxon rank-sum tests; outliers; large sample distribution; astronomical data; compatibility test.}
\section{Introduction}
Astronomical data, coming from different sources and collected by different telescopes, are often needed to be combined in a complete data set for study. In this situation, it is always very important to test compatibility of two data sets, collected in different surveys or measured with different resolutions, before pooling them together and they can only be combined when they are compatible (see, for example, De et al., 2014; Modak et al., 2017). That means, they should have approximately the same amount of observational error on an average. One possible way to deal with this situation is to carry out the following hypothesis testing problem under multivariate set up.

Let $(\mathbf{X_1,...,X_{n_1}})$ and $(\mathbf{Y_1,...,Y_{n_2}})$ be two independent samples from $p-$variate $(p\geq 2)$ populations with continuous distribution functions (d.f.'s) $F$ and $G$ respectively, where $G(\mathbf{x})=F(\mathbf{x}-\boldsymbol{\triangle})$ for all $\mathbf{x}\in R^p$. We consider the problem of testing the null hypothesis $H_0:\boldsymbol{\triangle}=\mathbf{0}$ against the alternative $H_1:\boldsymbol{\triangle}\neq\mathbf{0}$. Our test can be employed to solve the above stated problem and indicates compatibility only when the null hypothesis is accepted.

In this context, the Hotelling $T^2$ test ($HT$) is optimal and unbiased when $F$ is a $p-$variate normal d.f. However, for non-normal population, its finite sample unbiasedness is not certain (Seber, $2004$). It performs poorly for high-dimensional data (Bai and Saranadasa, $1996$) and the observations which are affected by outliers. Moreover, it is incomputable when $p>n_1+n_2-2$. In astronomy, data collection on celestial bodies is often obscured by bad weather conditions, obstruction by another celestial objects, instrumental restrictions, etc., and it cannot be repeated. So, we often get data which are contaminated with noise, affected by outliers or sparsely distributed (see, for example, Feigelson and Babu, 2013). In such situations, the asymptotic distribution of $HT$ based on approximating the population dispersion matrix by the sample dispersion matrix fails to attain the desired size of the test. Because the sample variance-covariance matrix is affected by the outliers and is not anymore a consistent estimator of its population version. This problem is resolved by considering rank tests based on mutual distances of observations in each component (see, for example, Jure\v{c}kov\'{a} and Kalina, $2012$; Marozzi, $2015, 2016$). In the present work, we apply this concept on Wilcoxon rank-sum statistic obtained from mutual distances between a first sample observation and the other observations in each component, and find the maximum of all the componentwise Wilcoxon rank-sum statistics. Thus we obtain $n_1$ such statistics for each of the first sample observations and combine them in an appropriate manner (see, for example, Jure\v{c}kov\'{a} and Kalina, $2012$) to define the ultimate test statistic. Our simulation study shows significant improvement in terms of the efficacy of the test, which is measured by empirical power.

Missions like Galaxy Evolution Explorer, Kepler Space Telescope, Hubble Space Telescope collect terabytes of astronomical data, which are preserved in virtual archives like Sloan Digital Sky Survey,
Multi-mission Archive at STSCI, NASA Extragalactic Data
base, Chandra (see, for example, Chattopadhyay and Chattopadhyay, 2014). They provide multivariate data sets of considerably large sizes. So our aim is to develop a test which is distribution-free asymptotically under certain conditions and we concentrate on the situations where $p<<n_1+n_2$ (see, for example, De et al., 2014; Modak et al., 2017). Simulation study shows that the power of the proposed test is strictly increasing in the sample sizes and/or in the number of components, and emphasizes the usability of the test in checking compatibility of two multivariate large sample astronomical data sets. Moreover, our test, as it is based on ranks, performs robustly in the presence of outliers. We consider two competitors, viz., the Wilcoxon rank-sum  based test in Jure\v{c}kov\'{a} and Kalina ($2012$), abbreviated as the $JK$ test and its analogous test, using $L_1-$norm instead of $L_2-$norm, abbreviated as the $JK_a$ test.

The paper is organized as follows. The proposed test and its properties are discussed in Section $2$. Section $3$ contains simulation study. Application to astronomical data sets is given in Section $4$. Section $5$ concludes.
\section{Test}
Here $\mathbf{X_i}=(X_{i1},X_{i2},...,X_{ip})',i=1,2,...,n_1$ and $\mathbf{Y_i}=(Y_{i1},Y_{i2},...,Y_{ip})',i=1,2,...,n_2$ are two samples, and $Z=(\mathbf{X_1},\mathbf{X_2},...,\mathbf{X_{n_1}},\mathbf{Y{_1}},\mathbf{Y{_2}},...,\mathbf{Y_{n_2}})'$ is the pooled sample of size $N=n_1+n_2$, in which $\mathbf{Z_j}=(Z_{1j},Z_{2j},...,Z_{Nj})'$ represents the $j-$th column of $Z,j=1,2,...,p$. Let $R_{ij}$ be the rank of $Z_{ij}$ in ${Z_{1j},Z_{2j},...,Z_{Nj}}, j=1,2,...,p$. Then, the componentwise Wilcoxon rank-sum statistics are
\begin{equation*}
W_j=\sum\limits_{i=n_1+1}^{N}R_{ij},j=1,2,...,p.
\end{equation*}
The corresponding rank matrix is $R=\big(R_{ij}\big)$, in which each of the $p$ columns represents a permutation of $\{1,2,...,N\}$. The matrix $R$ can be transformed to a matrix $R^*$ by permuting the rows of $R$ so that the first column of $R^*$ becomes in the natural order from $1$ to $N$. Let the conditional probability distribution of $R$ under $H_0$, given $R^*$, be $P$. Then $R$ has uniform distribution over $N!$ permutations of $\{1,2,...,N\}$ under $P$. Next, we consider
\begin{equation*}
W^o_j=\frac{W_j-E(W_j\vert P)}{\sqrt{V(W_j\vert P)}},
\end{equation*}
where $E(W_j\vert P)=\frac{n_2}{2}(N+1)$ and $V(W_j\vert P)=\frac{n_1n_2}{12}(N+1), j=1,2,...,p$.
Then, $\mathbf{W^o}=(W^o_1,W^o_2,...,W^o_p)'$ has zero mean vector and covariance matrix $H$ with elements $h_{jj'}=1$ for $j=j'$ and
\begin{equation*}
h_{jj'}=\frac{12}{N(N^2-1)}\sum_{i=1}^{N}\bigg(R_{ij}-\frac{N+1}{2}\bigg)\bigg(R_{ij'}-\frac{N+1}{2}\bigg)\hspace{.1 in} for \hspace{.1 in}j\neq j'.
\end{equation*}
Let us now assume that for each $N$, there are $n_1=n_1(N)$ and $n_2=n_2(N)$ such that $n_1+n_2=N$ and as $N\rightarrow \infty$,
\begin{equation}\label{largesample}
n_2\rightarrow \infty\hspace{.1 in}\text{but} \hspace{.1 in} \frac{n_2}{N}\rightarrow\lambda, 0<\lambda<1.
\end{equation}
Then, given $P$, the distribution of $\mathbf{W^o}$ under $H_0$ converges to $N_p(\mathbf{0},\Gamma)$, where the elements of $\Gamma$ are $\gamma_{jj'}=1$ for $j=j'$ and $\gamma_{jj'}=$ the grade correlation coefficient between component $j$ and component $j'$ for $j\neq j'$ (see, for example,  Chatterjee and Sen, $1964$; Puri and Sen, $1971$). Thus, defining the statistic
\begin{equation*}
T=\max\{W^o_j,1\leq j \leq p\},
\end{equation*}
the conditional distribution of $T$ given $P$ can be computed empirically under $H_0$ when $\Gamma$ is known. In practice, $\Gamma$ is unknown and is replaced by its consistent estimator $H$. Note that this criterion based on $T$ is equivalent to the Tippet criterion (see, Tippett, $1931$), where the minimum of the $p-$values over several tests is considered, which is also used as $TIP_{JK}$ in Marrozi ($2016$).

Here, given $X_{ij}$ $(i=1,2,...,n_1,j=1,2,...,p)$, we consider the distances
\begin{equation*}
d_{j}(i,i')=\vert X_{ij}-Z_{i'j}\vert,i'(\neq i)=1,2,...,N
\end{equation*}
from which we get $R_{j}(i,i')$ as the rank of $d_{j}(i,i')$ in the pooled sample. Then we find the componentwise Wilcoxon rank-sum statistics
\begin{equation*}
W_j(i)=\sum\limits_{l=n_1+1}^{N}R_{j}(i,l),j=1,2,...,p.
\end{equation*}
Note that $\{d_{j}(i,l),l(\neq i)=1,2,...,n_1\}$ represents sample corresponding to a distribution function $F_{ij}(x\vert X_{ij})$, whereas $\{d_{j}(i,l),l=n_1+1,n_1+2,...,N\}$ represents another sample corresponding to a distribution function $G_{ij}(x\vert X_{ij})$. So, we can frame $n_1$ testing problems in which the null and the corresponding alternative hypotheses are, respectively, $H_{0j}:F_{ij}(x\vert X_{ij})$ and $G_{ij}(x\vert X_{ij})$ are identical and $H_{1j}:G_{ij}(x\vert X_{ij})$ is stochastically larger than $F_{ij}(x\vert X_{ij})$ $,i=1,2,...,n_1$. Then, for each $i$, $H_0$ is true if $H_{0j}$ holds for all $j$ and $H_0$ is false if $H_{0j}$ does not hold for at least one $j$ $(i=1,2,...,n_1)$. To test these, assuming continuity of the distribution functions, we proceed in the following way.

Let for each $i$ and given reduced rank collection matrix, $P(i)$ be the conditional probability distribution of the rank collection matrix under $H_0,i=1,2,...,n_1$. Then $P(i)$'s are equiprobable on $(N-1)!$ permutations of $\{1,2,...,N-1\}$, and hence we compute
\begin{equation*}
W^{o}_j(i)=\bigg(W_j(i)-\frac{E\big(W_j(i)\vert P(i)\big)}{\sqrt{V\big(W_j(i)\vert P(i)\big)}}\bigg), j=1,2,...,p.
\end{equation*}
Let $r(i)$ denote the Spearman's rank correlation matrix for given $i$. Then, under \eqref{largesample} and given $\big(r(i),P(i)\big)$, the conditional distribution of $T(i)=\max\{W^{o}_j(i),1\leq j\leq p\}$ is asymptotically distribution-free. Let $\tilde{T}$ be such that for any $i$ $(=1,2,...,n_1)$,
\begin{equation*}
P\bigg(\tilde{T}=T(i)\bigg)=\frac{1}{n_1},
\end{equation*}
under both $H_0$ and $H_1$.
Here we also have random matrix $\tilde{r}$ and conditional probability distribution $\tilde{P}$ over the same probability measure space as that of $P(i)$ such that
\begin{equation}\label{given}
P\bigg(\tilde{r}=r(i)\bigg)=\frac{1}{n_1} \hspace{.1 in}\text{and}  \hspace{.1 in}P\bigg(\tilde{P}=P(i)\bigg)=\frac{1}{n_1}
\end{equation}
for $i=1,2,...,n_1$.
Hence, \eqref{given} implies
\begin{equation*}
P\bigg(\tilde{T}=T(i)\vert \tilde{P}=P(i), \tilde{r}=r(i)\bigg)=\frac{1}{n_1}, i=1,2,...,n_1.
\end{equation*}
Then $\tilde{T}$ can be taken as our test statistic, and our level $\alpha$ $(0<\alpha<1)$ critical region would be $\textbf{C}=\{\tilde{T}>\tilde{t}_\alpha\}$, where
\begin{equation}
P\bigg(\tilde{T}>\tilde{t}_\alpha\vert \tilde{P},\tilde{r}\bigg)=\frac{1}{n_1}\sum\limits_{i=1}^{n_1}P\bigg(T(i)>\tilde{t}_\alpha\vert P(i), r(i)\bigg)=\alpha.
\end{equation}

\section{Simulation study}
We draw two independent random samples of sizes $n_1$ and $n_2$ from $p-$variate $(p\geq 2)$ d.f.'s in which marginal distributions are $(i)$ $N(0,1)$ (Normal), $(ii)$ $C(0,1)$ (Cauchy), and $(iii)$ $LN(0,2.5)$ (Log-Normal).
Dependence among the components can be described by the following models:\\
$(a)$ Independence among the components of the parent d.f.\\
$(b)$ $t$ copula (Nelsen, $2006$) is an elliptical copula corresponding to a multivariate $t$ distribution wherein the Sklar's theorem establishes dependence structure among the components. Let $\Psi$ be the d.f. of the $p-$variate $t$ distribution and $\Psi_i$ be the d.f. of the $i-$th component with inverse function $\Psi_i^{-1},i = 1,2, . . . , p$. Then $t$ copula determined by $\Psi$ is
\begin{equation*}
C(u_1, . . . , u_p) = \Psi\{\Psi_1^{-1} (u_1), . . . , \Psi_p^{-1}(u_p)\}, 0<u_1,u_2,...,u_p<1.
\end{equation*}
We consider $t$ copula corresponding to the $p-$variate $t$ distribution with $2$ degrees of freedom and correlation matrix with all the off diagonal elements $0.15$.\\
$(c)$ Frank copula (Nelsen, $2006$) is an Archimedean copula established using the generator
\begin{equation*}
\phi=-ln\frac{e^{-\beta t}-1}{e^{-\beta}-1}
\end{equation*}
with $\beta\geq0$ in which larger $\beta$ indicates stronger dependence. The corresponding inverse function $\phi^{-1}$  is as follows
\begin{equation*}
C(u_1, . . . , u_p) = \phi^{-1} \{\phi(u_1) + . . . + \phi(u_p)\}, 0<u_1,u_2,...,u_p<1.
\end{equation*}
In particular, we choose Frank copula parameterized by $\beta=0.90$.

Location shift, denoted by $\boldsymbol{\mu}$, is added to the second sample and we consider the same shift $(\mu)$ to each component as all the marginal distributions are the same. Hence we compute size and power of the tests by taking $\boldsymbol{\mu}=\mathbf{0}$ and $\boldsymbol{\mu}\neq\mathbf{0}$, respectively. We take the nominal level $\alpha=0.05$ and set $n_1=n_2=50$ and $100$, as we consider the large sample situation, and $p=2,4,10$, as we are concerned with the situation $p<<N$. For normal parent the exact Snedecor's $F$ distribution of $HT$ is used, while the asymptotic $\chi^2$ distribution is applied to non-normal parents. Asymptotic normality for $JK$ and $JK_a$ is adopted and asymptotic distribution for $\tilde{T}$ is computed empirically. The outcomes are computed using $10000$ replications.

In simulation study, we estimate the error as follows (see, Marrozi, $2016$). Let $RH_i$ be the random variable denoting the rejection of $H_0$ in the $i-$th replication of the simulation and the probability of rejecting $H_0$ is $PRH$, then $RH_i\overset{iid}\sim Ber(1,PRH)$, $i=1,2,\ldots,REP$, where $REP$ is the number of replications. The power function is estimated from
\begin{equation*}
\sum\limits_{i=1}^{REP}RH_{i}\sim bin(REP,PRH).
\end{equation*}
So, the error in simulation can be estimated as the standard deviation of the estimated power function given by
\begin{equation*}
ER=\sqrt{\frac{PRH(1-PRH)}{REP}}.
\end{equation*}Now, for fixed $REP$, $ER$ is increasing in $PRH$ for $0<PRH<0.5$ and decreasing in $PRH$ for $0.5<PRH\leq1$. The maximum of $ER$ is attained at $PRH=0.5$, in which $ER$ is $\frac{0.5}{\sqrt{REP}}$. However, under $H_0$, i.e. when $PRH=\alpha$ is assumed, $ER=\sqrt{\frac{\alpha(1-\alpha)}{REP}}$. In our case, $REP=10000$ and $\alpha=0.05$ so that $ER\leq 0.005$, and under $H_0$, $ER=0.00218$.

We provide the empirical power study for the tests (see, Tables \ref{t1}--\ref{t3}).
All the tests satisfy the desired size condition except $HT$, which fails to attain the nominal level for non-normal distributions because of the effect of outliers in the distributions, and hence its power should not be taken under consideration. In all the situations, power of $\tilde{T}$ always increases in the sample sizes and/ or in the number of components. The same happens for all the tests in distribution $(i)$ (see, Table \ref{t1}). As expected, being the optimal test, $HT$ outperforms the others for normal d.f. $\tilde{T}$ is the second best test, except being slightly outperformed by $JK$ for $\mu=0.2$ with $n_1=n_2=50$ and $p=2,4$ under model $(b)$, and by $JK, JK_a$ for $\mu=0.2$ with $n_1=n_2=50$ and $p=10$ under model $(c)$. These exceptions can be ignored as computational error, however, when $n_1=n_2=100$, $\tilde{T}$ prominently shows its superiority over $JK$ and $JK_a$.

For normal distribution, $\tilde{T}$ has considerable high power for large sample sizes with increasing $p$. Showing similar performances in terms of power, $JK$ generally performs little better than $JK_a$ (except $\mu=1, n_1=n_2=100, p=10,$ model $(c)$). Powers of both $JK$ and $JK_a$ are increasing in $p$.

For distribution $(ii)$, Table \ref{t2} shows that $\tilde{T}$ significantly dominates $JK$ and $JK_a$. Power of $JK$ decreases in $p$ under model $(a)$ and model $(c)$. For model $(b)$, the power of $JK$ decreases when $\mu=0.5,1$, but increases when $\mu=2$. This indicates dependence of the power function on relative change in both $\mu$ and $p$. Power of $JK_a$ decreases in $p$ under model $(a)$ except when $\mu=2$ with $n_1=n_2=100$, which implies dependence of the power function on the values of $\mu,n_1,n_2$ and $p$. For model $(b)$ with $n_1=n_2=50$, power of $JK_a$ decreases in $p$ when $\mu=0.5,1$ and increases when $\mu=2$. Under the same model with $n_1=n_2=100$, power of $JK$ increases in $p$ when $\mu=1,2$ except when $\mu=0.5$. Under model $(c)$, power of $JK_a$ decreases in $p$ everywhere except when $\mu=2,n_1=n_2=100$.

Also, for distribution $(iii)$, $\tilde{T}$ has the best performance in terms of power for all the situations considered (Table \ref{t3}). Power of $JK_a$ decreases in $p$ under models $(a),(b)$ and also under model $(c)$ (except for $\mu=4$ with $n_1=n_2=100$). For $JK$, the increase in $p$ takes its toll on the decrease of power so much that slight change in the location shift cannot show up the change in power function correctly. For that reason, despite $JK$ being an unbiased test, the estimated power gets less than the estimated size for $p>2$. As a side effect the estimated power is a little less for $n_1=n_2=100$ than that for $n_1=n_2=50$ in model $(a)$ with $p=10, \mu=0.5$ and models $(b)$ and $(c)$ with $p=4, 10, \mu=0.5$. This problem can affect the application of the test to multivariate data with $p>2$, while our test being maximized over components does not suffer from such problems.

Now, in real life situations, the data sets may have different sizes, so we study the performance of our proposed test with significantly different sample sizes $n_1=50,n_2=100$ and $p=2,4,10$ (as considered before). Since we have seen (see, Tables \ref{t1}--\ref{t3}) that the tests' relative performance remains the same under all the three models, we consider the effect of unequal sample sizes under model $(a)$ (see, Table \ref{t4}). In Table \ref{t4}, we also study the empirical power for unequal location shifts in different components. Under all the situations, Table \ref{t4} shows that the relative performances of the tests in terms of power remains the same. Since the power of $\tilde{T}$ is strictly increasing in the sample sizes, we observe that the power for $n_1=50,n_2=100$ lies between those for $n_1=50,n_2=50$ and for $n_1=100,n_2=100$. Also, it is strictly increasing in $p$, provided the total of the shifts differs. Otherwise the power increases with the average of the shifts.
\section{Application}
We have $n_1$ dependent $p$-values,
\begin{equation}
p_i=P\bigg(T(i)>t_i\vert P(i), r(i)\bigg),
\end{equation}
of $T(i)$'s, where $t_i$ is the observed value for $T(i), i=1,2,...,n_1$. To take decision by the test we need to combine these $p$-values. There are various methods in the literature to combine independent $p$-values (see, for example, Tippett, 1931; Fisher, 1948; Liptak, 1958) and dependent $p$-values (see, for example, Brown, 1975; Kost and McDermott, 2002; Poole et al., 2016). We propose empirically to compute the lower $\alpha$ point, say $p_{\alpha}$ for $\sum\limits_{i=1}^{n_1}p_i$ and reject $H_0$ when $\sum\limits_{i=1}^{n_1}p_i<p_{\alpha}$. Here, $p_{\alpha}$ is computed using $B$ bootstrap values on $\sum\limits_{i=1}^{n_1}p_i$, say $\sum\limits_{i=1}^{n_1}p_{i,b},b=1,2,...,B$ where $p_{i,b}$ is computed, for each $i$, from $p-$variate two-sample bootstrap samples on interpoint distances $,b=1,2,...,B$. Similarly, we compute the lower $\alpha$ points for the total of the $p$-values corresponding to $JK$ and $JK_a$ by performing univariate two-sample bootstrapping on interpoint distances for each $i$.

\begin{figure}
\centering
\includegraphics[width=1\textwidth]{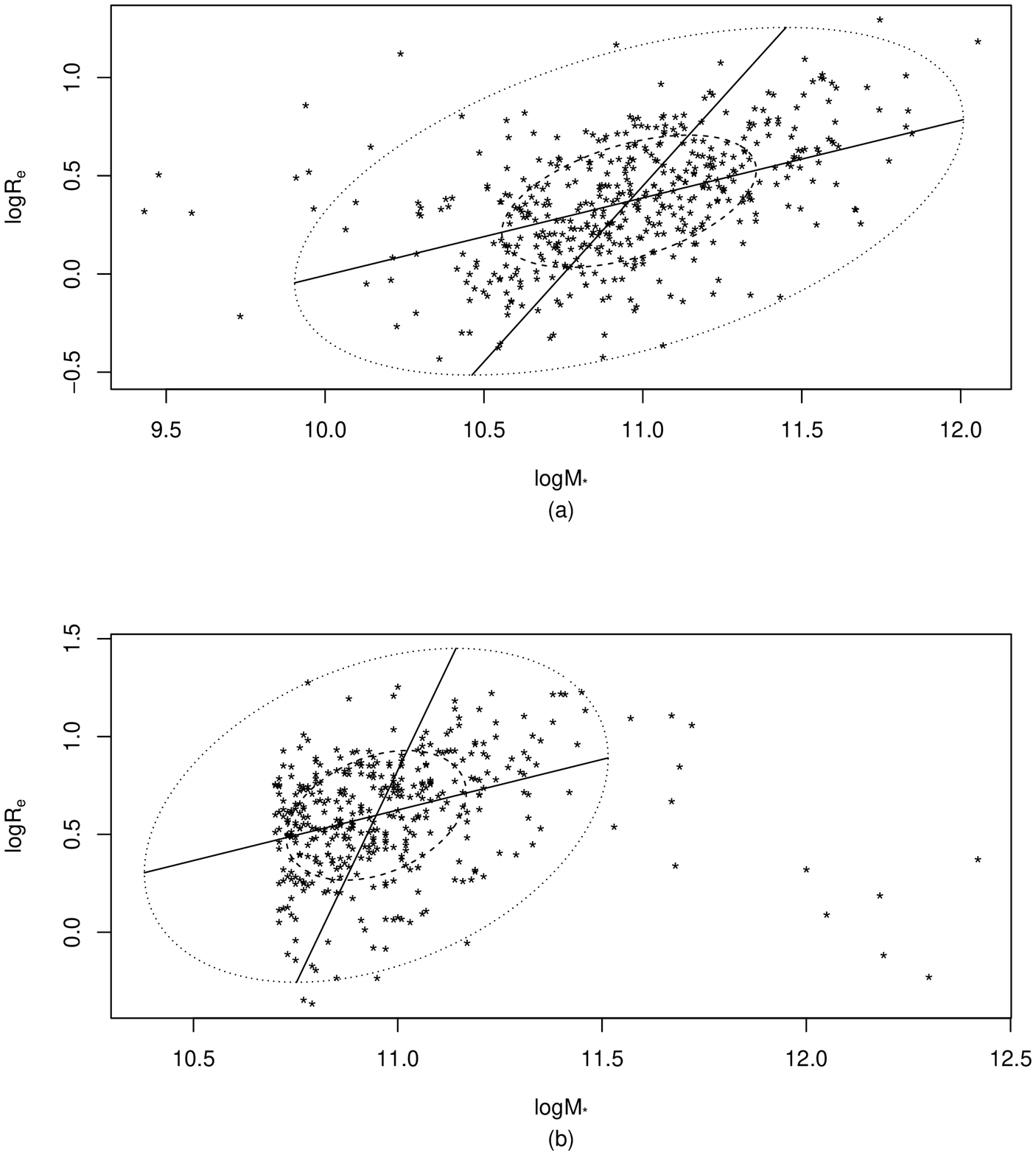}
\caption{Bivariate boxplot on the mass ($logM_{*}$) versus size ($logR_{e}$) parameter space for $(a)$ the first data set and $(b)$ the second data set, where the observations lying outside the outer ellipse indicate the potential outliers.}
\label{f1}
\end{figure}

An important area of study in astronomy is the formation and evolution of a certain kind of galaxies called early-type galaxies (ETGs) (De et al., $2014$; Modak et al., 2017). Here, data are collected from different sources and therefore needed to be checked for compatibility before pooling them together for further study. So, we perform compatibility test between the first data set containing $465$ ETGs in the redshift range $0.2<z<2.7$, collected from Damjanov et al. $2011$, and the second data set consisting $397$ ETGs in the redshift range $0<z<2.5$, collected from Szomoru et al. $2013$, on mass-size parameter space. We draw a bivariate boxplot (Fig. \ref{f1}) using robust biweight M estimators of correlation,
scale and location (Goldberg and Iglewicz, $1992$; for practical implementation see, Everitt, $2005$) on mass-size observations. Fig. \ref{f1} shows that both the data sets are affected by outliers, which encourages us to apply our test to these data sets of quite large sizes. The $p$-value based on $HT$ is $0$; and the totals of the $p$-values for the other tests, with the lower $\alpha$ $(=0.05)$ points for $B=500$, are computed as: (test, $\sum\limits_{i=1}^{n_1}p_i,p_{\alpha})=(JK, 233.11, 229.34), (JK_a, 234.71, 231.92)$ and $(\tilde{T}, 158.71, 149.37)$. As it is always computationally convincing to take the data set of smaller size as the first data set, we perform the following tests interchanging the data sets and get (test, $\sum\limits_{i=1}^{n_1}p_i,p_{\alpha})=(JK, 74.10, 71.87), (JK_a, 70.40, 67.16)$ and $(\tilde{T}, 6.25, 4.39)$. As $HT$ is supposed to be misleading in such situation, we can ignore it; and the $p-$values of the other tests support compatibility of the two data sets.
\section{Conclusion}
We propose a nonparametric test using Wilcoxon rank-sum test statistics on distances between observations for each of the components. The test is asymptotically distribution-free under certain conditions. The simulation study shows that the test is unbiased and its power is strictly increasing in the sample sizes and/or in the number of components, provided $p<<N$, which encourages its applicability to multivariate large sample astronomical data sets. In the presence of outliers or sparsely distributed data where $HT$ fails, the performance of our proposed test,
measured in terms of power, is the best among the possible competitors. For distribution $(i)$, $HT$ is optimal but under all the models power of $\tilde{T}$ becomes very close to $1$ for $n_1=n_2=100$ with $p=10$. It guarantees its good performance for the parent distributions like multivariate normal when the sample sizes are large. It is to be noted that in greater effect of outliers as in distribution $(iii)$ than in distribution $(ii)$, $\tilde{T}$ performs better than $JK$, $JK_a$ with higher efficacy. It indicates the proposed test's robustness under the presence of unusual observations in the parent distributions. $JK$ and $JK_a$ not only get outperformed in the above stated situations but also their powers may become significantly worse for increasing values of $p$. While our test being maximized over the
components gets only better with increasing $p$, provided $p<<N$. Unlike $HT$, our test is computable for $p>N$, but not suitable for use, since the test depends on the central limit theorem, which does not hold for large $p$. As our objective is to provide a test for large sample data with $p<<N$, we do not concern this problem here, while our future project is to provide tests for large-dimensional data.

\clearpage

\begin{table}
\caption{Simulation study for distribution $(i)$}
\begin{center}
\tiny
\begin{tabular}{*{9}{c}}
\hline\hline
\noalign{\vskip .05in}
$\mu$ &  $HT$  & $JK$ & $JK_a$ & $\tilde{T}$ &  $HT$  & $JK$ & $JK_a$ & $\tilde{T}$\\
[1ex]
   && $n_1=n_2=50$&&                    &&$n_1=n_2=100$&&  \\

&&&&model $(a)$&&&&\\
\hline
&&&&$p=2$&&&&\\

$0$  &0.04990& 0.04861& 0.04867& 0.05026&0.05020& 0.04918& 0.04925& 0.04927\\
$0.2$ &0.20830& 0.10459& 0.10255& 0.10579&0.40810& 0.15008& 0.14610& 0.16179\\
$0.5$ &0.88610& 0.33980& 0.32535& 0.40829&0.99600& 0.43237& 0.41226& 0.57800\\
$1$  &1.00000& 0.62325& 0.60290& 0.77786&1.00000& 0.66822& 0.65175& 0.82652\\
[1ex]

&&&& $p=4$&&&&\\

$0$&0.04720& 0.04941& 0.04907& 0.05008&0.05130& 0.05122& 0.05122& 0.05201\\
$0.2$&0.29860& 0.11518& 0.11049& 0.12142&0.58670& 0.16841& 0.15968& 0.20597\\
$0.5$&0.98580& 0.38583& 0.36446& 0.54071&1.00000& 0.47835& 0.45632& 0.77191\\
$1$&1.00000& 0.71481& 0.69811& 0.93690&1.00000& 0.75848& 0.74582& 0.96507\\
[1ex]

&&&&$p=10$&&&&\\

$0$&0.04800& 0.04912& 0.04929& 0.04869&0.05090& 0.05079& 0.05089& 0.05059\\
$0.2$&0.48890& 0.13195& 0.12450& 0.14676&0.86820& 0.19375& 0.18180& 0.26499\\
$0.5$&1.00000& 0.48031& 0.45679& 0.73100&1.00000& 0.57846& 0.55675& 0.94623\\
$1$&1.00000& 0.86505& 0.85299& 0.99798&1.00000& 0.89469& 0.88598& 0.99962\\
[1ex]

&&&&model $(b)$&&&&\\
\hline
&&&&$p=2$&&&&\\
$0$&0.05040& 0.04855& 0.04889& 0.05000&0.04820& 0.04825& 0.04805& 0.04892\\
$0.2$&0.20000& 0.10831& 0.10497& 0.10829&0.36130& 0.15603& 0.15033& 0.16172\\
$0.5$&0.83390& 0.34474& 0.32599& 0.39441&0.99000& 0.43772& 0.41201& 0.55802\\
$1$&1.00000& 0.63498& 0.60622& 0.75302&1.00000& 0.68464& 0.65820& 0.80255\\
[1ex]

&&&& $p=4$&&&&\\
$0$&0.04740& 0.05138& 0.05067& 0.04759&0.04870& 0.05032& 0.05020& 0.04975\\
$0.2$&0.22160& 0.12275& 0.11635& 0.11936&0.43570& 0.17853& 0.16779& 0.19616\\
$0.5$&0.92370& 0.39789& 0.37163& 0.50403&0.99910& 0.49066& 0.46296& 0.71665\\
$1$&1.00000& 0.71459& 0.68479& 0.89962&1.00000& 0.75597& 0.73027& 0.93481\\
[1ex]

&&&& $p=10$&&&&\\
$0$&0.04380& 0.04934& 0.04916& 0.04787&0.04630& 0.04840& 0.04852& 0.04710\\
$0.2$&0.22750& 0.13985& 0.13257& 0.14116&0.46470& 0.20313& 0.19146& 0.24185\\
$0.5$&0.96180& 0.46113& 0.42923& 0.64367&1.00000& 0.54796& 0.51730& 0.86595\\
$1$&1.00000& 0.79265& 0.76156& 0.98162&1.00000& 0.82218& 0.79450& 0.99284\\
[1ex]

&&&&model $(c)$&&&&\\
\hline
&&&&$p=2$&&&&\\

$0$&0.04830& 0.05005& 0.05009& 0.05029&0.05280& 0.05012& 0.04970& 0.04858\\
$0.2$&0.20090& 0.10386& 0.10292& 0.10897&0.36470& 0.15136& 0.14926& 0.16556\\
$0.5$&0.83600& 0.33174& 0.32418& 0.39563&0.99110& 0.43115& 0.41928& 0.56336\\
$1$&1.00000& 0.60766& 0.59553& 0.75024&1.00000& 0.65282& 0.64315& 0.80190\\
[1ex]

&&&&$p=4$&&&&\\
$0$&0.05120& 0.04809& 0.04805& 0.04729&0.04660& 0.04945& 0.04934& 0.04857\\
$0.2$&0.21980& 0.11513& 0.11308& 0.11861&0.42770& 0.17059& 0.16693& 0.19493\\
$0.5$&0.92450& 0.38401& 0.37548& 0.50380&0.99910& 0.47840& 0.47105& 0.71400\\
$1$&1.00000& 0.67680& 0.67227& 0.88490&1.00000& 0.71647& 0.71441& 0.92272\\
[1ex]

&&&&$p=10$&&&&\\
$0$&0.04670& 0.04886& 0.04985& 0.04820&0.05050& 0.05066& 0.05048& 0.04856\\
$0.2$&0.21170& 0.14447& 0.14159& 0.13956&0.44310& 0.21871& 0.21347& 0.24267\\
$0.5$&0.95160& 0.48744& 0.48102& 0.64578&1.00000& 0.57285& 0.56997& 0.85207\\
$1$&1.00000& 0.76157& 0.76164& 0.96055&1.00000& 0.78689& 0.78787& 0.97899\\
\hline
\end{tabular}
\end{center}
\label{t1}
\end{table}

\clearpage
\begin{table}
\caption{Simulation study for distribution $(ii)$}
\begin{center}
\tiny
\begin{tabular}{*{9}{c}}
\hline\hline
\noalign{\vskip .05in}
$\mu$ &  $HT$  & $JK$ & $JK_a$ & $\tilde{T}$ &  $HT$  & $JK$ & $JK_a$ & $\tilde{T}$\\
[1ex]
   && $n_1=n_2=50$&&                    &&$n_1=n_2=100$&&  \\

&&&&model $(a)$&&&&\\
\hline
&&&&$p=2$&&&&\\
$0$&0.02020& 0.04986& 0.05003& 0.05162&0.02220& 0.05062& 0.05111& 0.04930\\
$0.5$&0.04020& 0.12425& 0.11972& 0.20081&0.04100& 0.18545& 0.17507& 0.32703\\
$1$&0.10080& 0.30662& 0.29661& 0.53505&0.10440& 0.42408& 0.40246& 0.70580\\
$2$&0.31850& 0.60763& 0.59517& 0.86324&0.31430& 0.66951& 0.65146& 0.89987\\
[1ex]

&&&& $p=4$&&&&\\
$0$&0.02150& 0.04942& 0.04953& 0.04926&0.01850& 0.05162& 0.05145& 0.04924\\
$0.5$&0.04590& 0.09344& 0.09182& 0.24182&0.04310& 0.13244& 0.12773& 0.42802\\
$1$&0.13900& 0.21634& 0.22583& 0.69362&0.13210& 0.32036& 0.33015& 0.88324\\
$2$&0.47260& 0.51314& 0.57641& 0.97484&0.46810& 0.61502& 0.67826& 0.98837\\
[1ex]

&&&&$p=10$&&&&\\
$0$&0.04620& 0.04980& 0.05070& 0.04888&0.02350& 0.05075& 0.05095& 0.04985\\
$0.5$&0.09530& 0.07073& 0.07591& 0.31715&0.06390& 0.08809& 0.09370& 0.58087\\
$1$&0.28720& 0.13119& 0.16971& 0.87569&0.23480& 0.18997& 0.25268& 0.98861\\
$2$&0.78360& 0.33411& 0.55579& 0.99971&0.76080& 0.46128& 0.71071& 0.99998\\
[1ex]

&&&&model $(b)$&&&&\\
\hline
&&&&$p=2$&&&&\\
$0$&0.02050& 0.04925& 0.04927& 0.04956&0.01900& 0.04751& 0.04760& 0.04856\\
$0.5$&0.04360& 0.14418& 0.14119& 0.19176&0.04040& 0.21590& 0.20879& 0.31494\\
$1$&0.11070& 0.35666& 0.34585& 0.51250&0.10550& 0.46983& 0.44473& 0.68350\\
$2$&0.33490& 0.65410& 0.63319& 0.83699&0.31770& 0.70366& 0.67867& 0.87410\\
[1ex]

&&&& $p=4$&&&&\\
$0$&0.02330& 0.05186& 0.05199& 0.04832&0.01670& 0.05066& 0.05063& 0.05001\\
$0.5$&0.05810& 0.13533& 0.13746& 0.23075&0.04870& 0.19977& 0.20167& 0.39721\\
$1$&0.17190& 0.33425& 0.34683& 0.64052&0.16190& 0.45054& 0.45939& 0.83744\\
$2$&0.51190& 0.65713& 0.68077& 0.94620&0.50640& 0.71942& 0.73728& 0.96716\\
[1ex]

&&&& $p=10$&&&&\\
$0$&0.04500& 0.05039& 0.04974& 0.04778&0.02820& 0.04851& 0.04816& 0.04796\\
$0.5$&0.12310& 0.13194& 0.14209& 0.28418&0.09120& 0.19001& 0.20510& 0.50170\\
$1$&0.38870& 0.32605& 0.36955& 0.77558&0.33770& 0.43870& 0.48551& 0.94679\\
$2$&0.86540& 0.67241& 0.73565& 0.99257&0.84400& 0.74121& 0.78818& 0.99756\\
[1ex]

&&&&model $(c)$&&&&\\
\hline
&&&&$p=2$&&&&\\
$0$&0.02010& 0.04952& 0.04969& 0.05095&0.01890& 0.05033& 0.05030& 0.04914\\
$0.5$&0.04130& 0.12200& 0.11887& 0.19491&0.03550& 0.18568& 0.17946& 0.32062\\
$1$&0.09660& 0.30300& 0.29908& 0.51542&0.09370& 0.42455& 0.41264& 0.68493\\
$2$&0.30470& 0.60143& 0.59657& 0.83968&0.30430& 0.66317& 0.65190& 0.87976\\
[1ex]

&&&&$p=4$&&&&\\
$0$&0.02420& 0.04657& 0.04631& 0.04733&0.01910& 0.04868& 0.04924& 0.04888\\
$0.5$&0.04700& 0.09268& 0.09497& 0.23300&0.04080& 0.13522& 0.13638& 0.40541\\
$1$&0.13530& 0.22143& 0.24235& 0.65384&0.12430& 0.32766& 0.35357& 0.83165\\
$2$&0.44290& 0.50875& 0.57972& 0.94188&0.44490& 0.60340& 0.66993& 0.96518\\
[1ex]

&&&&$p=10$&&&&\\
$0$&0.04340& 0.04996& 0.04908& 0.04637&0.02450& 0.04888& 0.04891& 0.04840\\
$0.5$&0.08160& 0.07495& 0.08250& 0.29625&0.05670& 0.09325& 0.10739& 0.53160\\
$1$&0.22550& 0.14349& 0.19841& 0.80607&0.19150& 0.21027& 0.29842& 0.93235\\
$2$&0.69770& 0.35168& 0.56579& 0.98614&0.68740& 0.47222& 0.68718& 0.99393\\

\hline
\end{tabular}
\end{center}
\label{t2}
\end{table}

\clearpage
\begin{table}
\caption{Simulation study for distribution $(iii)$}
\begin{center}
\tiny
\begin{tabular}{*{9}{c}}
\hline\hline
\noalign{\vskip .05in}
$\mu$ &  $HT$  & $JK$ & $JK_a$ & $\tilde{T}$ &  $HT$  & $JK$ & $JK_a$ & $\tilde{T}$\\
[1ex]
   && $n_1=n_2=50$&&                    &&$n_1=n_2=100$&&  \\

&&&&model $(a)$&&&&\\
\hline
&&&&$p=2$&&&&\\
$0$&0.01670& 0.04860& 0.04913& 0.05007&0.01820& 0.05213& 0.05197& 0.05098\\
$0.5$&0.01930& 0.05010& 0.07260& 0.29478&0.01960& 0.06794& 0.09912& 0.44568\\
$1$&0.02300& 0.12009& 0.16155& 0.55875&0.01990& 0.17847& 0.22896&0.66438\\
$2$&0.03410& 0.28643& 0.33552& 0.76093&0.03250& 0.36887& 0.40546& 0.78773\\
[1ex]

&&&& $p=4$&&&&\\
$0$&0.02100& 0.04892& 0.04848& 0.04883&0.01920& 0.05046& 0.05023& 0.05023\\
$0.5$&0.02120& 0.03113& 0.05331& 0.40191&0.02020& 0.03119& 0.06254& 0.63403\\
$1$&0.02470& 0.03848& 0.08644& 0.75341&0.02450& 0.04747& 0.12061& 0.87745\\
$2$&0.03740& 0.08498& 0.20992& 0.93304&0.03670& 0.12969& 0.31388& 0.95345\\
[1ex]

&&&&$p=10$&&&&\\
$0$&0.04150& 0.04855& 0.04797& 0.04935&0.02940& 0.05060& 0.05109& 0.05026\\
$0.5$&0.04430& 0.03744& 0.05214& 0.57740&0.02620& 0.03695& 0.05836& 0.86089\\
$1$&0.05130& 0.03328& 0.06638& 0.93581&0.03040& 0.03330& 0.08220& 0.99213\\
$2$&0.08200& 0.03452& 0.12425& 0.99787&0.05250& 0.03931& 0.18156& 0.99949\\
[1ex]

&&&&model $(b)$&&&&\\
\hline
&&&&$p=2$&&&&\\
$0$&0.01530& 0.04711& 0.04664& 0.04933&0.01900& 0.04818& 0.04867& 0.04935\\
$0.5$&0.01640& 0.05478& 0.07683& 0.28757&0.01920& 0.07935& 0.11138& 0.43066\\
$1$&0.02060& 0.13982& 0.18040& 0.54179&0.02120& 0.20920& 0.25594& 0.64266\\
$2$&0.03200& 0.32038& 0.36491& 0.73876&0.03070& 0.39742& 0.42623& 0.76371\\
[1ex]

&&&& $p=4$&&&&\\
$0$&0.02000& 0.05273& 0.05296& 0.04756&0.02010& 0.05022& 0.05022& 0.04999\\
$0.5$&0.02040& 0.03554& 0.06363& 0.38177&0.02080& 0.03431& 0.07472& 0.58822\\
$1$&0.02340& 0.05582& 0.12196& 0.70939&0.02470& 0.07387& 0.17742& 0.83032\\
$2$&0.04060& 0.14748& 0.30428& 0.89858&0.03430& 0.21902& 0.41903& 0.91995\\
[1ex]

&&&& $p=10$&&&&\\
$0$&0.03990& 0.04994& 0.04979& 0.04865&0.01980& 0.04924& 0.04892& 0.04766\\
$0.5$&0.04090& 0.03292& 0.05842& 0.51813&0.02170& 0.02987& 0.06500& 0.77113\\
$1$&0.04870& 0.03554& 0.09940& 0.87520&0.02560& 0.03860& 0.13762& 0.96254\\
$2$&0.08070& 0.06947& 0.25620& 0.98456&0.04210& 0.09769& 0.37465& 0.99164\\
[1ex]

&&&&model $(c)$&&&&\\
\hline
&&&&$p=2$&&&&\\
$0$&0.01910& 0.05027& 0.05037& 0.04929&0.01910& 0.05059& 0.05055& 0.04982\\
$0.5$&0.01990& 0.05894& 0.08176& 0.28633&0.01950& 0.08246& 0.11418& 0.43444\\
$1$&0.02320& 0.14019& 0.17977& 0.53668&0.02250& 0.21135& 0.25605& 0.64127\\
$2$&0.03550& 0.30849& 0.35384& 0.73412&0.03410& 0.39298& 0.42336& 0.76108\\
$4$&0.08720& 0.50450& 0.52960& 0.84125&0.07770& 0.53416& 0.55572& 0.84650\\
[1ex]

&&&&$p=4$&&&&\\
$0$&0.01850& 0.05042& 0.05035& 0.04810&0.01760& 0.04752& 0.04739& 0.04920\\
$0.5$&0.01900& 0.03139& 0.05668& 0.37887&0.01900& 0.02918& 0.06327& 0.58760\\
$1$&0.02220& 0.04546& 0.10240& 0.69800&0.02180& 0.05729& 0.14449& 0.81464\\
$2$&0.03990& 0.10956& 0.24807& 0.88368&0.03260& 0.16678& 0.36307& 0.90440\\
$4$&0.11500& 0.28597& 0.52206& 0.94971&0.09320& 0.37953& 0.62186& 0.95303\\
[1ex]

&&&&$p=10$&&&&\\
$0$&0.04260& 0.04832& 0.04929& 0.04850&0.02450& 0.05178& 0.05136& 0.04870\\
$0.5$&0.04500& 0.03364& 0.05212& 0.50428&0.02600& 0.03320& 0.05676& 0.75702\\
$1$&0.04980& 0.03065& 0.07013& 0.84828&0.03050& 0.03102& 0.08848& 0.92942\\
$2$&0.07590& 0.03641& 0.14514& 0.96485&0.04950& 0.04349& 0.22206& 0.97423\\
$4$&0.21250& 0.07776& 0.38907& 0.99073&0.15020& 0.11836& 0.55619& 0.99163\\
\hline
\end{tabular}
\end{center}
\label{t3}
\end{table}
\clearpage
\begin{table}
\caption{Simulation study with $n_1=50,n_2=100$ under model $(a)$}
\begin{center}
\tiny
\begin{tabular}{*{5}{c}}
\hline\hline
\noalign{\vskip .05in}
$\boldsymbol{\mu}$ &  $HT$  & $JK$ & $JK_a$ & $\tilde{T}$\\
[1ex]

&&distribution $(i)$&&\\
[1ex]
\hline

&&$p=2$&&\\
$(0,0)'$&0.04820& 0.05083& 0.05093& 0.05093\\
$(0.2,0)'$&0.15340& 0.08480& 0.08349& 0.08843\\
$(0.2,0.2)'$&0.28250& 0.11935& 0.11682& 0.12711\\
$(0.5,0.5)'$&0.95740& 0.37824& 0.36185& 0.48012\\
$(0.2,1)'$&1.00000& 0.52272& 0.52739& 0.58506\\
$(1,1)'$&1.00000& 0.64485& 0.62597& 0.80268\\
[1ex]
&&$p=4$&&\\
0$\boldsymbol{\epsilon}_4'$& 0.04450& 0.04865& 0.04940& 0.04942\\
(0.2,0$\boldsymbol{\epsilon}_3)'$& 0.12610& 0.07106& 0.06958& 0.07732\\
0.2$\boldsymbol{\epsilon}_4'$& 0.40820& 0.13507& 0.12796& 0.14970\\
0.5$\boldsymbol{\epsilon}_4'$&0.99810& 0.42643& 0.40400& 0.64296\\
(0.2$\boldsymbol{\epsilon}_2$,$\boldsymbol{\epsilon}_2)'$&1.00000& 0.59068& 0.57347& 0.79896\\
$\boldsymbol{\epsilon}_4'$&1.00000& 0.73537& 0.72004& 0.95277\\
[1ex]
&&$p=10$&&\\
0$\boldsymbol{\epsilon}_{10}'$&0.04810& 0.04989& 0.04974& 0.04779\\
(0.2,0$\boldsymbol{\epsilon}_9)'$&0.08950& 0.05847& 0.05734& 0.06150\\
0.2$\boldsymbol{\epsilon}_{10}'$&0.66580& 0.15341& 0.14377& 0.18195\\
0.5$\boldsymbol{\epsilon}_{10}'$&1.00000& 0.52358& 0.50011& 0.84373\\
(0.2$\boldsymbol{\epsilon}_5$,$\boldsymbol{\epsilon}_5)'$& 1.00000& 0.72707& 0.70878& 0.97304\\
$\boldsymbol{\epsilon}_{10}'$&1.00000& 0.87911& 0.86862& 0.99912\\
[1ex]

&&distribution $(ii)$&&\\
[1ex]
\hline
&&$p=2$&&\\
$(0,0)'$&0.02940& 0.05068& 0.05089& 0.05124\\
$(0.5,0)'$&0.03680& 0.09680& 0.09338& 0.15283\\
$(0.5,0.5)'$&0.04260& 0.14302& 0.13628& 0.24141\\
$(1,1)'$&0.09570& 0.35198& 0.33706& 0.60960\\
$(0.5,2)'$& 0.16910& 0.48841& 0.52513& 0.70064\\
$(2,2)'$& 0.28880& 0.63499& 0.61981& 0.88128\\
[1ex]

&&$p=4$&&\\
0$\boldsymbol{\epsilon}_4'$&0.02780& 0.05036& 0.05005& 0.04889\\
(0.5,0$\boldsymbol{\epsilon}_3)'$& 0.03510& 0.06509& 0.06413& 0.12687\\
0.5$\boldsymbol{\epsilon}_4'$&0.04870& 0.10642& 0.10339& 0.31446\\
$\boldsymbol{\epsilon}_4'$& 0.12720& 0.26088& 0.27029& 0.79132\\
(0.5$\boldsymbol{\epsilon}_2$,2$\boldsymbol{\epsilon}_2)'$& 0.24940& 0.38700& 0.44951& 0.89102\\
2$\boldsymbol{\epsilon}_4'$&0.43520& 0.55409& 0.61734& 0.98224\\
[1ex]
&&$p=10$&&\\
0$\boldsymbol{\epsilon}_{10}'$& 0.04580& 0.04860& 0.04831& 0.04957\\
(0.5,0$\boldsymbol{\epsilon}_9)'$&0.05340& 0.05090& 0.05060& 0.09559\\
0.5$\boldsymbol{\epsilon}_{10}'$&0.08220& 0.07624& 0.08176& 0.42333\\
$\boldsymbol{\epsilon}_{10}'$&0.23590& 0.15501& 0.20153& 0.94782\\
(0.5$\boldsymbol{\epsilon}_5$,2$\boldsymbol{\epsilon}_5)'$&0.44730& 0.23820& 0.37550& 0.99295\\
2$\boldsymbol{\epsilon}_{10}'$&0.72340& 0.38738& 0.62087& 0.99992\\
[1ex]
&&distribution $(iii)$&&\\
[1ex]
\hline
&&$p=2$&&\\
$(0,0)'$&0.02880& 0.05029& 0.05022& 0.04956\\
$(0.5,0)'$& 0.02930& 0.05739& 0.07498& 0.21654\\
$(0.5,0.5)'$&0.02440& 0.05731& 0.08356& 0.35175\\
$(1,1)'$&0.02740& 0.14573& 0.18898& 0.60333\\
$(.5,2)'$&0.03120& 0.24685& 0.34408& 0.61864\\
$(2,2)'$&0.03280& 0.31736& 0.36483& 0.77470\\
[1ex]
&&$p=4$&&\\
0$\boldsymbol{\epsilon}_4'$&0.03060& 0.05215& 0.05256& 0.05122\\
(0.5,0$\boldsymbol{\epsilon}_3)'$&0.02950& 0.04410& 0.05104& 0.18859\\
0.5$\boldsymbol{\epsilon}_4'$&0.02820& 0.03332& 0.06094& 0.50360\\
$\boldsymbol{\epsilon}_4'$& 0.02550& 0.04135& 0.10046& 0.81013\\
(0.5$\boldsymbol{\epsilon}_2$,2$\boldsymbol{\epsilon}_2)'$&0.03010& 0.07293& 0.16318& 0.83274\\
2$\boldsymbol{\epsilon}_4'$&0.03380& 0.10350& 0.24930& 0.94364\\
[1ex]
&&$p=10$&&\\
0$\boldsymbol{\epsilon}_{10}'$&0.05100& 0.05170& 0.05153& 0.04850\\
(0.5,0$\boldsymbol{\epsilon}_9)'$& 0.04710& 0.04837& 0.05015& 0.15525\\
0.5$\boldsymbol{\epsilon}_{10}'$&0.04080& 0.03536& 0.05252& 0.71634\\
$\boldsymbol{\epsilon}_{10}'$& 0.04150& 0.03214& 0.07388& 0.97144\\
(0.5$\boldsymbol{\epsilon}_5$,2$\boldsymbol{\epsilon}_5)'$&0.05210& 0.03680& 0.09647& 0.98267\\
2$\boldsymbol{\epsilon}_{10}'$&0.05210& 0.03511& 0.14717& 0.99895\\
[2ex]
\hline
\noalign{\vskip .05in}
Note: $\boldsymbol{\epsilon}_i$ is an $i\times1$\\
vector with all entries\\
equal to 1, $i\geq2$.
\end{tabular}
\end{center}
\label{t4}
\end{table}
\clearpage
{}
\end{document}